\documentclass[12pt]{article}
\usepackage{amsmath,amssymb,amsxtra}
\usepackage[dvips]{graphicx}

\def\pzp{p_0^{\prime}}

\newcommand{\nbp}{|\mbox{\boldmath $p$}|}

\newcommand{\bp}{{\bf p}}
\newcommand{\bpp}{{\bf p}^{\prime}}
\newcommand{\bk}{{\bf k}}

\newcommand{\thpp}{\theta_{\mbox{\boldmath ${\bf{\hat p}}$}}}
\newcommand{\phpp}{\varphi_{\mbox{\boldmath ${\bf{\hat p}}$}}}
\newcommand{\bhp}{{\bf{\hat p}}}

\begin{document}

\begin{center}
{\bfseries RELATIVISTIC CONTRIBUTION OF THE FINAL-STATE\\
INTERACTION TO DEUTERON PHOTODISINTEGRATION}

\vskip 5mm

S.G. Bondarenko$^{a,}$\footnote{E-mail: bondarenko@jinr.ru},
V.V. Burov$^{a,}$\footnote{E-mail: burov@thsun1.jinr.ru},
K.Yu. Kazakov$^{b,}$\footnote{E-mail: kazakovk@ifit.phys.dvgu.ru},
D.V. Shulga$^{b,}$\footnote{E-mail: denis@ifit.phys.dvgu.ru}

\vskip 5mm

{\small
$^a$ {\it
Joint Institute for Nuclear Research, 141980 Dubna, Russia
}
\\
$^b$ {\it
Far Eastern National University, Sukhanov Str. 8,\\
Vladivostok, 690600, Russia
}}
\end{center}

\vskip 5mm

\begin{center}
\begin{minipage}{150mm}
\centerline{\bf Abstract} The contribution of the final-state
interaction to the differential cross-section of deuteron
photodisintegration at laboratory photon energies below 50 MeV is
analyzed in the framework of Bethe--Salpeter formalism with a
phenomenological rank-one separable interaction. The
approximations made are the neglect of two-body exchange currents,
negative-energy components of the bound-state vertex function and
the scattering $T$ matrix. It has been found that the gross effect
of the final-state interaction with $J\leqslant 1$ comes from the
net contributions of the spin-triplet final states and
spin-singlet $^1S_0^{+}$-state. The relativistic results are
compared with the nonrelativistic ones in every partial-wave
channel. It is found that the relativistic effects
change the magnitude of the final-state interaction
from several percent to several tens of percent.
\end{minipage}
\end{center}

\vskip 10mm

\section*{Introduction}\indent

The investigation of deuteron photodisintegration is a direct way
of probing the structure of light nuclei. High-precision
experimental data on polarization observables of the reaction can
give information about the nucleon-nucleon ($NN$) interaction. On
the one hand, the important issue related to the nuclear force is
the relativistic effects, which may play an important role in the
nuclear two-body problem (the deuteron-bound state and the
final-state interaction (FSI) of the final neutron-proton ($np$)
state). On the other hand, in deuteron photodisintegration, the
electromagnetic (EM) form factors of nucleons are probed as well.
Knowledge of the deuteron-bound state, the final-state interaction
in two-nucleon scattering states and two-body currents can help to
estimate the half on-mass-shell behavior of the nucleon EM form
factors.

The present work is devoted to a comprehensive study of
contributions to the differential cross section caused by the
relativistic structure of $np$ scattering states. These are
analyzed in terms of the field-theoretical nonperturbative theory,
based on the Bethe-Salpeter (BS) equation. The development of
previous works~\cite{KS,KS1} is necessary to test the
sensitivity of observables to the FSI due to the relativistic
nuclear force and relativistic effects of kinematical origin. To
this end, we employ a separable interaction kernel of the
$NN$ interaction to solve the BS equation for two Dirac particles in
Minkowski space. As a result, we obtain the relativistic deuteron
vertex function in the coupled $^3S_1$--${~}^3D_1$ channel and the
scattering $T$ matrix of the elastic $NN$ scattering for the
positive-energy partial-wave channels with the total angular
momentum $J=0,1$ at laboratory energies
$T_{\text{lab}} \leqslant 100$~MeV~\cite{PPNP,NPA}.

This paper is organized as follows: in Sec. 1, we present the
basic information on the kinematics of the reaction, the invariant
transition amplitude and the differential cross-section. The
relativistic approach to the $NN$ interaction is briefly described
in Sec. 2. The rank-one separable interaction kernel is
introduced in Sec. 3. In Sec. 4, we go into several points of
calculating the invariant transition amplitudes within the BS
formalism. Discussions of principle results and final remarks are
in Sec. 5.

\section{Kinematics of the reaction}\indent

If we denote by $|F(P_f,p_f,\zeta_f)\rangle$ any particular
neutron-proton final state, the invariant transition amplitude
$\mathcal M_{fi}$ for deuteron photodisintegration in the c.m.
frame being the rest frame of the $np$ pair
\begin{equation}\label{deuteron breakup}
\gamma(q,\varepsilon)+d(K_i,\xi_i)\rightarrow F(P_f,p_f,\zeta_{f})
\end{equation}
is given by the matrix element  of the hadronic EM four current
operator ${\widehat J^{\mu}}(x)$ between the initial deuteron
state and final $np$ pair state
\begin{equation}\label{transition amplitude}
\mathcal M_{fi} = \varepsilon^\mu(\lambda)\cdot\langle
F(P_f,p_f,\zeta_f) | {\widehat J_{\mu}}(0)|
d(K_i,\xi_i(m_d))\rangle,
\end{equation}
where $P_f = (\sqrt{s},{\bf 0})$ and $p_f = (0,\bhp)$ are
total and asymptotic relative four momenta of the pair, $K_i$ and
$q=P_f-K_i$ are the deuteron and photon four momenta,
respectively. Four polarization four vectors
$\varepsilon(\lambda)$, $\xi_i(m_d)$ and $\zeta_f$ (with
$\lambda=\pm1$ and $m_d=0,\pm1$) describe the internal degrees of
freedom of the photon, deuteron and final $np$ system, related
to the angular momentum. All states in Eq.~\eqref{transition
amplitude} are understood to be normalized in the covariant
manner. It should be also noted that the definition of the
outgoing $np$ pair is such that whether the final-state
interaction is switched off, it is given by a antisymmetrized
product of two positive-energy Dirac spinors.

The coordinate system is defined by the incoming three momentum of
the photon, $\mathbf q =(0,0,\omega)$ with $|\mathbf q|
=\omega$, which is along the $z$ axis, and transverse
polarizations
$\boldsymbol{\varepsilon}_{\lambda=\pm 1}=(\mp 1,- i,0)/\sqrt{2}$.
The asymptotic
relative three momentum ${\bf\hat p}$ is characterized by the
spherical angles $\thpp$ and $\phpp$ in the chosen coordinate
system. The $z$ axis is also the quantization axis for the total
spin $S=0,1$ of the final $np$ system, as well as for the
polarization four-vector $\xi_i(m_d)$ of the deuteron.

The c.m. energy squared $s$ of the final $np$ pair is related to
the relative three momentum $\hat {\mathbf p}$, $s = 4({\hat {\bf p}}^2 +
m^2)$, and to the photon energy $E_{\gamma}$ in the laboratory
system, being the rest frame of the deuteron,
$s=M_{d}^2+2E_{\gamma}M_{d}$, where $m$ and $M_{d}$ are the
nucleon and deuteron masses, respectively. The nucleon laboratory
energy $T_{\text{lab}}=\tfrac{2}{m}\bhp^2$ is connected to the
laboratory photon energy as $E_\gamma\cong\tfrac12T_{\text{lab}}$.

The differential cross section, in the case of the unpolarized
initial configuration, can be written in the c.m. frame as follows
\begin{eqnarray}
\frac{d\sigma}{d\Omega_{\bhp}}=
\dfrac{\alpha}{16\pi s} \dfrac{|\bhp|}{\omega}\,\overline{\mid {\cal
M}_{fi}\mid^{2}}, \label{c00}
\end{eqnarray}
where $\alpha={e^2}/{4\pi}$ is the fine structure constant
and line over $\mid {\cal M}_{fi}\mid^{2}$ is the incoherent
summation over $\lambda$, $m_d$
and $m_s$, being the spin projection of the final $np$ system in
spin-triplet channels, on the quantization axis.

We assume in this work that the dynamical model of the EM current
operator in Eq.~\eqref{transition amplitude} is the relativistic
impulse approximation not constrained by gauge invariance. For
simplicity, we do not modify this approximation by applying
Siegert's theorem. This is a major drawback of the present
development, preventing us from comparing of theoretical
predictions of the relativistic theory and experimental results
for deuteron photodisintegration at low energies.

All details concerning the structure of the EM current operator
and the initial deuteron state could be found in
Ref.~\cite{KS,KS1}, where numerical calculations of the angular
distribution in the plane wave approximation (PWA) have been done.
Thus, in determining the differential cross section~\eqref{c00},
we are focused on the relativistic structure of the final
two-nucleon system.

\section{The two-body problem in momentum space}\indent

Within the relativistic field theory, the elastic $NN$ scattering
can be described by the scattering $T$ matrix, which satisfies the
inhomogeneous BS equation. In momentum space, the BS equation for
the $T$ matrix reads (in terms of the relative four momenta
$p^\prime$ and $p$ and the total four momentum $P_f$)
\begin{eqnarray}
T(p^{\prime}, p; P_f) = V(p^{\prime}, p; P_f) + \frac{i}{4\pi^3}\int
d^4k\, V(p^{\prime}, k; P_f)\, S_2(k; P_f)\, T(k, p; P_f), \label{t00}
\end{eqnarray}
where $V(p^{\prime}, p; P_f)$ is the interaction kernel and
$S_2(k; P_f)$ is free two-particle Green's function
$$S_2^{-1}(k; P_f)=\bigl(\tfrac12\:P_f\cdot\gamma+{k\cdot\gamma}-m\bigr)^{(1)}
\bigl(\tfrac12\:P_f\cdot\gamma-{k\cdot\gamma}-m\bigr)^{(2)}.$$

To perform the partial-wave decomposition of the BS equation (\ref{t00})
we introduce relativistic two-nucleon basis states
$|aM\rangle\equiv|\pi,\,{}^{2S+1}L_J^{\rho} M \rangle$, where $S$
denotes the total spin, $L$ is the orbital angular moment and $J$
is the total angular momentum with the projection $M$;
relativistic quantum numbers $\rho$ and $\pi$ refer to the total
energy-spin and relative-energy parity
with respect to the change of sign of the
relative energy, respectively. Then the partial-wave decomposition
of the $T$ matrix in the c.m. frame has the following form
\begin{eqnarray}
T_{\alpha\beta,\gamma\delta}(p^{\prime},p; {P_{f(0)}}) = \sum_{JMab}
({\cal Y}_{aM}(-{\bpp})U_C)_{\alpha\beta}\otimes (U_C {\cal
Y}^{\dag}_{bM}({\bp}))_{\delta\gamma}\
t_{ab}(\pzp,|\bpp|;p_0,|\bp|;s), \label{t01}
\end{eqnarray}
where $U_C =i\gamma^2\gamma^0$ is the charge conjugation matrix.
Greek letters $(\alpha,\beta)$ and $(\gamma,\delta)$ in
Eq.~\eqref{t01} refer to spinor indices and label particles in the
initial and final states, respectively. It is convenient to
represent the two-particle states in terms of matrices. To this
end the Dirac spinors of the second nucleon are transposed. At
this stage $T$ is $16\times16$ matrix in spinor space which,
sandwiched between Dirac spinors and traced, yields the
corresponding transition matrix elements between $SLJ$-states.

The spin-angular momentum functions ${\cal Y}_{aM}({\bp})$ are
expressed in terms of the positive- and negative-energy Dirac
spinors $u^{\rho=\pm 1/2}_{m}$, the spherical harmonics
$Y_{L{m_L}}$ and Clebsch-Gordan coefficients $C_{j_1 m_1 j_2
m_2}^{j\:m}$
\begin{eqnarray} {\cal Y}_{JM:LS {\rho}}(\bp) U_C
\hskip 110mm\nonumber\\
= i^{L} \sum_{m_Lm_Sm_1m_2\rho_1\rho_2} C_{\frac12 \rho_1 \frac12
\rho_2}^{S_{\rho} {\rho}} C_{L m_L S m_S}^{JM} C_{\frac12 m_1
\frac12 m_2}^{Sm_S} Y_{L{m_L}}(\bp) {u^{\rho_1}_{m_1}}^{(1)}(\bp)
{{u^{\rho_2}_{m_2}}^{(2)}}^{T}(-\bp). \label{s01}
\end{eqnarray}
The superscripts in Eq.~\eqref{s01} refer to particles (1) and (2).
In deriving the matrix elements between $a$-states, the
ortonormalization condition for the functions ${\cal
Y}_{aM}({\bpp})$ should be used
\begin{eqnarray}
\int\!d\varphi_{\bp}\:d(\cos\theta_{\bp})\,{\rm Tr}\left\{ {\cal
Y}^{\dag}_{aM}({\bp}){\cal
Y}_{a^{\prime}M^{\prime}}({\bp})\right\}
\hskip 80mm\nonumber\\
\equiv \int\!d\varphi_{\bp}\:d(\cos\theta_{\bp}) ({\cal
Y}^{\dag}_{aM}({\bp}))_{\beta\alpha} ({\cal
Y}_{a^{\prime}M^{\prime}} ({\bp}))_{\alpha\beta} =
\delta_{aa^{\prime}}\delta_{MM^{\prime}}, \label{t02}
\end{eqnarray}
where partial states $a$ and $a^{\prime}$ belong to the same partial channel.

The partial-wave decomposition for the interaction kernel $V$ of
the BS equation~\eqref{t00}  can be written analogously to
Eq.~\eqref{t01}
\begin{eqnarray}
V_{\alpha\beta,\gamma\delta}(p^{\prime},p; P_{f(0)}) = \sum_{abM}
({\cal Y}_{aM}(-{\bpp})U_C)_{\alpha\beta}\otimes (U_C{\cal
Y}^{\dag}_{bM}({\bp}))_{\delta\gamma}\
v_{ab}(\pzp,|\bpp|;p_0,|\bp|;s). \label{t03}
\end{eqnarray}
Applying the condition~\eqref{t02}, we can obtain a system of
linear  integral equations of the off-shell partial-wave
amplitudes
\begin{eqnarray}
t_{ab}(\pzp, |\bpp|; p_0, |\bp|; s) =
v_{ab}(\pzp, |\bpp|; p_0, |\bp|; s)
\hskip 70mm
\label{t04}\\
+ \frac{i}{4\pi^3}\sum_{cd}\int\limits_{-\infty}^{+\infty}\!
dk_0\int\limits_0^\infty\! \bk^2 d|\bk|\, v_{ac}(\pzp, |\bpp|; k_0,
|\bk|; s)\, S_{cd}(k_0,|\bk|;s)\, t_{db}(k_0,|\bk|;p_0,|\bp|;s), \nonumber
\end{eqnarray}
where the two-particle propagator $S_{ab}$ depends only on
$\rho$-spin indices.

The PWA and FSI contributions can be combined by introducing the
relativistic scattering amplitude for two nucleons
$\chi_{Sm_s}(p;p_fP_f)$, which satisfies the following equation
\begin{eqnarray}
\chi_{Sm_s}(p;p_f,P_f) = \chi_{Sm_s}^{(0)}(p;p_f,P_f) +
\dfrac{i}{4\pi^3} S_2(p;P_f)\int\!d^4k\,
V(p,k;P_f)\chi_{Sm_s}(k;p_f,P_f), \label{t05c}
\end{eqnarray}
with $p_f\cdot P_f = 0$ and $p_f^2 = -s/4+m^2$ putting the outgoing
particles onto the mass shell. The first term
$\chi_{Sm_s}^{(0)}(p;p_f,P_f)$ in Eq.~\eqref{t05c} is the PWA
amplitude, which describes the free motion of two nucleons. Due to
Pauli's principle, it is the antisymmetric combination of
positive-energy Dirac spinors and isovector, isocalar factors.
Omitting isospin we can write in c.m. frame
\begin{eqnarray}
\chi_{Sm_s}^{(0)}(p;p_f,P_{f(0)}) = \delta^{(4)}(p-p_f)
\chi_{Sm_s}^{(0)}(p_f,P_{f(0)})
\hskip 60mm\nonumber\\
= \delta^{(4)}(p-p_f) \sum_{m_1m_2} C_{\frac12 m_1 \frac12
m_2}^{Sm_S} u_{m_1}^{(1)}\left(\hat{\mathbf p}\right)
u_{m_2}^{(2)}\left(-\hat{\mathbf p}\right).
\end{eqnarray}
It is easy to rewrite the FSI contribution to the BS amplitude,
the second term in Eq.~\eqref{t05c}, from the $T$ matrix, once the
following relation is used,
\begin{eqnarray}
\int\! d^4k\,V(p,k;P_f)\chi_{Sm_s}(k;p_f,P_f) = \int\!
d^4k\,T(p,k;P_f)\chi_{Sm_s}^{(0)}(k;p_f,P_f).
\end{eqnarray}
The result is
\begin{equation}\label{}
\chi_{Sm_s}^{(t)}(p;p_f,P_f)=\dfrac{i}{4\pi^3}S_2(p;P_f)T(p,p_f;P_f)
\chi_{Sm_s}^{(0)}(p_f,P_{f}).
\end{equation}
The ultimate expression for the BS scattering amplitude of the
$np$ pair is solely defined by the $T$ matrix half on the mass-shell
\begin{eqnarray}
\chi_{Sm_s}(p;p_f,P_f) = \bigl[\delta^{(4)}(p-p_f)+
\dfrac{i}{4\pi^3}S_2(p;P_f)T(p,p_f;P_f)\bigr]
\chi_{Sm_s}^{(0)}(p_f,P_f). \label{t05b}
\end{eqnarray}
The partial-wave decomposition of $\chi_{Sm_s}^{(t)}(p;p_fP_f)$
can be written in the form
\begin{eqnarray}
\chi_{Sm_s}^{(t)}(p;p_f,P_{f(0)}) = \sum_{LmJMa} C_{LmSm_s}^{JM}\,
Y^{*}_{Lm}(\thpp,\phpp)\, {\cal
Y}_{aM}(\bp)\,\phi_{a,J:LS\rho=+1}(p_0,|\bp|;0,|\bhp|;s),
\label{t06}
\end{eqnarray}
where the radial function $\phi$ is determined by the product of
the transition matrix elements $t$ and the two-particle propagator
\begin{eqnarray}
\phi_{a,J:LS\rho=+1}(p_0,|\bp|;0,|\bhp|;s)= S_{+a}(p_0,|\bp|;s)\,
t_{a,J:LS\rho=+1}(p_0,|\bp|;0,|\bhp|;s). \label{t07}
\end{eqnarray}

The inclusion of intermediate partial-wave channels with the
negative-energy Dirac spinors leads to large sets of
two-dimensional equations in the system~\eqref{t04}. Its solution
determines the relativistic wave function, as well as the phase
shifts from the $T$ matrix. For simplicity, we switch off all
partial-wave channels with negative-energy states and focus only
on the physical channels with the total angular moment
$J\leqslant1$. Splitting the channels with $J=0$ and $J=1$, the
partial-wave decomposition for the conjugate BS amplitude has the
form
\begin{eqnarray}
{\bar \chi}_{Sm_s}^{(t)}(p;p_f,P_{f(0)})=
\sum\limits_{LmL^{\prime}S^{\prime}} C_{LmSM_S}^{00}\,
Y_{Lm}(\thpp,\phpp)\, {\bar {\cal
Y}}_{00:L^{\prime}S^{\prime}}(\bp)\,
\phi^*_{0:LS,0:L^{\prime}S^{\prime}}(0,|\bhp|;p_0,|\bp|;s)
\label{t06a}\\
+\sum\limits_{MLmL^{\prime}S^{\prime}} C_{LmSM_S}^{1M}\,
Y_{Lm}(\thpp,\phpp)\, {\bar {\cal
Y}}_{1M:L^{\prime}S^{\prime}}(\bp)\,
\phi^*_{1:LS,1:L^{\prime}S^{\prime}}(0,|\bhp|;p_0,|\bp|;s),
\nonumber\end{eqnarray} where the spin-angular momentum functions
${\bar {\cal Y}}_{aM}(\bp)$ for the positive-energy states can be
written in the matrix form
\begin{eqnarray}
{\bar {\cal Y}}_{aM}(\bp) = \frac{1}{\sqrt{8\pi}}\,
\frac{1}{4E_{\bp}(E_{\bp}+m)}\, (m-p_2\cdot\gamma)\, {\bar {\cal
G}}_{aM} (1+\gamma_0)\, (m+p_2\cdot\gamma), \label{t11}
\end{eqnarray}
where $E_{\bp}=\sqrt{m^2+\bp^2}$ and explicit expressions for the
matrices ${\bar {\cal G}}_{aM}(\bp)$ are given in Table~\ref{saf}.

Further, we explicitly factor in contributions of the spin-singlet
states $^1S_0^+$ and $^1P_1^+$, uncoupled spin-triplet states
$^3P_0^+$ and $^3P_1^+$, and the coupled spin-triplet states
$^3S_1^+$--$\:{}^3D_1^+$ in Eq.~\eqref{t06a}. The result is
presented in a lengthy form
\begin{eqnarray}
&&{\bar \chi}_{Sm_s}^{(t)}(p;p_f,P_f)
=\dfrac{i}{4\pi^3}S_{++}(p_0,|\bp|;s)
\\
&\times\Bigl[& \delta_{S0}\,\delta_{m_S0}\,
\frac{1}{\sqrt{4\pi}}\, {\bar {\cal Y}}_{^1S_0^+}(\bp)\,
t_{^1S_0^+,^1S_0^+}^*(0,|\bhp|;p_0,|\bp|;s)\,
\nonumber\\
&+& \delta_{S0}\,\delta_{m_S0}\, \sum_{M} Y_{1M}(\thpp,\phpp)\,
{\bar {\cal Y}}_{^1P_1^+ M}(\bp)\,
t_{^1P_1^+,^1P_1^+}^*(0,|\bhp|;p_0,|\bp|;s)\,
\nonumber\\
&+& \delta_{S1}\, (-)^{1-m_S}\,\frac{1}{\sqrt{3}}\,Y_{1
-m_S}(\thpp,\phpp)\, {\bar {\cal Y}}_{^3P_0^+}(\bp)\,
t_{^3P_0^+,^3P_0^+}^*(0,|\bhp|;p_0,|\bp|;s)\,
\nonumber\\
&+& \delta_{S1}\, \sum_{Mm} C_{1m1m_S}^{1M}\,Y_{1m}(\thpp,\phpp)\,
{\bar {\cal Y}}_{^3P_1^+ M}(\bp)\,
t_{^3P_1^+,^3P_1^+}^*(0,|\bhp|;p_0,|\bp|;s)\,
\nonumber\\
&+& \delta_{S1}\, \frac{1}{\sqrt{4\pi}}\, {\bar {\cal Y}}_{^3S_1^+
m_S}(\bp)\, t_{^3S_1^+,^3S_1^+}^*(0,|\bhp|;p_0,|\bp|;s)
\nonumber\\
&+& \delta_{S1}\, \sum_{M} C_{2M-m_S 1m_S}^{1M}\,Y_{2
M-m_S}(\thpp,\phpp)\, {\bar {\cal Y}}_{^3S_1^+ M}(\bp)\,
t_{^3S_1^+,^3D_1^+}^*(0,|\bhp|;p_0,|\bp|;s)\,
\nonumber\\
&+& \delta_{S1}\, \frac{1}{\sqrt{4\pi}}\, {\bar {\cal Y}}_{^3D_1^+
m_S}(\bp)\, t_{^3D_1^+,^3S_1^+}^*(0,\bhp;p_0,|\bp|;s)
\nonumber\\
&+& \delta_{S1}\, \sum_{M} C_{2M-m_S 1m_S}^{1M}\,Y_{2
M-m_S}(\thpp,\phpp)\, {\bar {\cal Y}}_{^3D_1^+ M}(\bp)\,
t_{^3D_1^+,^3D_1^+}^*(0,|\bhp|;p_0,\nbp;s)\, \Bigr], \nonumber
\end{eqnarray}
where
$S^{-1}_{++}(k_0,|\bk|;s)=\left[(\sqrt{s}/2-E_{\bk}+i0)^2-k_0^2\right]$
is the projection of the two-particle propagator onto positive-energy
states.

At this stage, the contribution of the FSI to the BS amplitude of
the $np$ pair is expressed in terms of the spin-angular momentum
functions and radial parts of the half off-mass-shell $T$ matrix in
$SLJ$-representation. To perform further calculations, we need to
solve the BS equation~\eqref{t04}.
\begin{table}
{\caption{\label{saf} Spin-angular momentum matrices $\bar {\cal
G}_{aM} $ for $J\leqslant 1$. \protect \\  $p_1=(E_{\bp},{\bp})$,
$p_2=(E_{\bp},-{\bp})$ are on-mass-shell four momenta.}}
\vskip 11pt
\begin{tabular}{cc}
\hline\hline\\
$^1S_0^+$ & $\gamma_5$ \\
$^3P_0^+$ & $-\tfrac{1}{2|\bp|}\:(p_1-p_2)\cdot\gamma$ \\
$^3S_1^+$ & $-\xi^*_f(M)$ \\
$^1P_1^+$ & $\tfrac{\sqrt{3}}{|\bp|}\:p\cdot\xi^*_f(M)\gamma_5$ \\
$^3P_1^+$ &
$\sqrt{\tfrac32}\,\gamma_5\bigl[\tfrac12(p_1-p_2)\cdot\gamma\:
\xi^*_f(M)\cdot\gamma-p\cdot\xi^*_f(M)\bigr]\tfrac{1}{|\bp|}$ \\
$^3D_1^+$ & $-\tfrac{1}{\sqrt{2}}
\bigl[\xi^*_f(M)\cdot\gamma+\tfrac32(p_1-p_2)\cdot\gamma\:p
\cdot\xi^*_f(M)\bigr]\tfrac{1}{|\bp|^2}$
\\\\
\hline\hline
\end{tabular}
\end{table}

\section{A separable kernel of the {\boldmath $NN$} interaction}\indent

First, we assume that the interaction kernel $V$ conserves parity,
total spin $S$, total angular momentum  $J$ and its projection,
and isotopic spin. Because of the tensor nuclear force, the
orbital angular momentum $L$ is not conserved. Moreover, the
negative-energy two-nucleon states are switched off. The
partial-wave-decomposed BS equation is therefore decomposed to the
following form
\begin{eqnarray}
t_{LL^{\prime}}(\pzp, |\bpp|; p_0, |\bp|; s) =
v_{LL^{\prime}}(\pzp, |\bpp|; p_0, |\bp|; s) \hskip 70mm
\label{t04b}\\
+ \frac{i}{4\pi^3}\sum_{L^{\prime\prime}}
\int\limits_{-\infty}^{+\infty}\!dk_0\int\limits_0^\infty\! \bk^2
d|\bk| v_{LL''}(\pzp, |\bpp|; k_0,|\bk|; s)
S_{++}(k_0,|\bk|;s) t_{L^{\prime\prime}L}(k_0,|\bk|;p_0,|\bp|;s),
\nonumber
\end{eqnarray}
where $L=L^{\prime}=J$ is for spin-singlet and uncoupled spin-triplet
states and $L,L^{\prime}=J\pm1$ is for coupled spin-triplet states.

Next, we make a rank-one separable \emph{ansatz} for the
$NN$ interaction kernel of the form
\begin{eqnarray}
v_{L^{\prime}L}(\pzp, |\bpp|; p_0, |\bp|; s) =
\lambda\,g^{(L^{\prime})}(\pzp, |\bpp|)\,
g^{(L)}(p_0, |\bp|),
\label{t18}\end{eqnarray} where $\lambda$ is the coupling
strength and $g^{(L)}(p_0, |\bp|)$ are the covariant form factors,
which depend only upon the zero components $p_0, p_0'$ and
magnitudes $|\bpp|,|\bp|$ of the spatial components of the relative
four momenta. In Eq.~\eqref{t18} the partial-wave channels for a
given $J$ are labeled only by the angular momenta $L,L'$.

Thus, the two-fold integrals in Eq.~\eqref{t04b} can be solved in
a closed form, yielding the final expression for the $T$ matrix
elements
\begin{eqnarray}
t_{L^{\prime}L}(\pzp, |\bpp|, p_0, |\bp|; s) = \tau(s)\,
g^{(L^{\prime})}(\pzp, |\bpp|) g^{(L)}(p_0, |\bp|),\label{t19}\end{eqnarray}
where function $\tau(s)$ has the form
\begin{equation}\label{}
\tau(s)=
\dfrac{1}{\lambda^{-1} + h(s)},
\end{equation}
with
\begin{eqnarray}
h(s) = -\frac{i}{4\pi^3}\,\sum_{L}
\int\limits_{-\infty}^{+\infty}\!dk_0\,\int\limits_0^\infty\!\bk^2\,
d|\bk|\,\dfrac{g^{(L)}(k_0,|\bk|)g^{(L)}(k_0,|\bk|)}
{(\sqrt{s}/2-E_{\bk}+i0)^2-k_0^2}.
\label{t20}\end{eqnarray}

The nuclear phase shifts $\delta(s)$ are related to the fully
on-mass-shell $T$ matrix through the condition of the two-body elastic
unitarity. For the spin-singlet and uncoupled spin-triplet states
the parameterizations of the on-mass-shell $T$ matrix have the form
\begin{eqnarray}
t_L(s) \equiv t_{LL}(0,|{\bf p}|; 0,|{\bf p}|;s) = -
\dfrac{16 \pi}{\sqrt{s}\sqrt{s-4m^2}}\, \text{e}^{i\delta_L(s)}\,
\sin{\delta_L(s)}, \label{t21}\end{eqnarray} where
$\delta_L(s)\equiv\delta_{L=J}(s)$

For the coupled spin-triplet states, we use the parameterization
\begin{equation}\label{}
t_{L^{\prime}L}(s)=
\frac{8\pi i}{\sqrt{s}\sqrt{s-4m^2}}
\left(
\begin{array}{cc}
\cos{2\epsilon_J}\ e^{2i\delta_{<}} - 1 &
i\sin{2\epsilon_J}\ e^{i(\delta_{<}+\delta_{>})} \\
i\sin{2\epsilon_J}\ e^{i(\delta_{<}+\delta_{>})} &
\cos{2\epsilon_J}\ e^{2i\delta_{>}} - 1\\
\end{array}
\right),
\end{equation}
in terms of the phase shifts
$\delta_{\lessgtr}\equiv\delta_{L=J\mp 1}$ and the mixing
parameters $\epsilon_J(s)$.

In numerical calculations of the phase shifts, we consider the
covariant relativistic generalizations of the {\em Yamaguchi}-type
form factors
\begin{align}\label{}
  g^{(L=0)}(k_0,|\bk|) & = \dfrac{1}{k_0^2-\bk^2-\beta_0^2+i0},\\
  g^{(L=1)}(k_0,|\bk|) & =
  \dfrac{\sqrt{|-k_0^2+\bk^2|}}{(k_0^2-\bk^2-\beta_1^2+i0)^{2}},\\
  g^{(L=2)}(k_0,|\bk|) & =
  \dfrac{C(k_0^2-\bk^2)}{(k_0^2-\bk^2-\beta_2^2+i0)^{2}}.
\end{align}

The nuclear phase shifts $\delta$, mixing parameters $\epsilon$,
as well as the low-energy $NN$ parameters (the scattering length and
the effective range) and the deuteron static properties (the
binding energy,  a $D$-state probability and the quadrupole
moment) can be computed in terms of internal parameters
$\lambda$, $C$ and $\beta_{0,1,2}$ through a specially
designed procedure. Technical details can be found in Ref~\cite{NPA}.
Values of the parameter are in Table~\ref{TabP1}.

\begin{table}
\caption{Kernel parameters for partial-wave channels with $J \leqslant 1$}
\label{TabP1}
\begin{tabular}{lccccc}
\hline
                    &$^1S_0^+$  &
                          $^3S_1^+-^3D_1^+$ &
                                              $^3P_0^+$ & $^1P_1^+$ & $^3P_1^+$  \\
\hline
$\lambda$ (GeV$^2$) & -0.28554  & -0.50269 & -0.016123  & 0.091535  & 0.084724   \\
$\beta_0$ (GeV)     &  0.221858 &  0.25124 &            &           &            \\
$\beta_1$ (GeV)     &           &          & 0.21861    & 0.27673   & 0.28398    \\
$\beta_2$ (GeV)     &           &  0.29399 &            &           &            \\
C                   &           &  1.6471  &            &           &            \\
\hline
\end{tabular}
\end{table}

\section{Calculation of the transition amplitudes}\indent

In the framework of the BS formalism, the invariant transition
amplitude, given by Eq.~\eqref{transition amplitude}, can be
written as
\begin{eqnarray}
\mathcal{M}_{fi}(q,K_i)=\int\!
d^4p\:d^4k\:\bar{\chi}_{Sm_S}(p;p_f,P_f)\:\varepsilon(\lambda)\cdot
\tilde{J}(p,k;P_f,K_i)~S_2(k,K_i)
\Gamma_{m_d}(k,K_i),
\label{amplitude in BS}\end{eqnarray}
where
subscripts $f$ and $i$ imply the polarization quantum numbers of
the final $(Sm_s)$ and initial $(\lambda\:m_d)$ states,
respectively; the BS amplitude $\bar{\chi}_{Sm_s}(p;p_f,P_f)$ of
the $np$ pair is given by Eq.~\eqref{t06a};
$\Gamma_{m_d}(k,K_i)$ is the deuteron vertex function, which is
the solution of the homogenous BS equation with the same
interaction kernel $V$; $\tilde{J}(p,k;P_f,K_i)$ is Mandelstam's
EM vertex operator, describing the interaction of the photon with
the hadronic system. The structure of the EM current operator is
specified by the microscopic model such as the impulse
approximation (IA).

The invariant transition amplitude of Eq.\eqref{amplitude in BS}
can be cast into the form
\begin{equation}\label{amplitude casting}
\mathcal{M}_{fi}=\mathcal{M}_{fi,IA}^{PWA}+\mathcal{M}_{fi,IA}^{FSI}+\mathcal{M}_{fi,TB},
\end{equation}
where $\mathcal{M}_{fi,IA}^{PWA}$ is the transition amplitude in
the plane-wave impulse approximation; $\mathcal{M}_{fi,IA}^{FSI}$
determines contributions of FSI to the impulse approximation, and
$\mathcal{M}_{fi,TB}$ accounts for two-body exchange currents. In
the present work we consider only the first two terms in
Eq.~\eqref{amplitude casting}, and their explicit expressions are
expressed by
\begin{eqnarray}
\mathcal{M}_{fi,IA}^{PWA}&=&-\varepsilon(\lambda)\,\text{Sp}\biggl\{\bar\chi^{(0)}_{Sm_S}(p_f,P_f)
\:\left[\Gamma^{(1)}\left(q\right)+(-1)^{S+1}
\Gamma^{(2)}\left(q\right)\right]\Lambda(\mathcal L)
\nonumber\\&&~~ \times\:S\left(\mathcal
L^{-1}(\tfrac12(K_i-q)+p_f)\right) \Gamma_{m_d}\left(\mathcal
L^{-1}(p_f-\tfrac12q);K_{i(0)}\right)\Lambda(\mathcal
L)^{-1}~\biggr\},\label{MPWA}
\\
\mathcal{M}_{fi,IA}^{FSI}&=&-\varepsilon(\lambda)\,\int\!
d^4p\,\text{
Sp}\biggl\{\bar{\chi}_{Sm_S}^{(t)}(p;p_f,P_f)\:\left[\Gamma^{(1)}\left(q\right)+(-1)^{S+1}
\Gamma^{(2)}\left(q\right)\right]\Lambda(\mathcal
L)\nonumber\\&&~~~~~~~~ \times S\left(\mathcal
L^{-1}(\tfrac12(K_i-q)+p\right) \Gamma_{m_d}\left(\mathcal
L^{-1}(p-\tfrac12q);K_{i(0)}\right)\Lambda(\mathcal
L)^{-1}\biggr\},\label{MFSI}
\end{eqnarray}
where $\Gamma^{(1,2)}$ refers to the $\gamma NN$ vertex operator of
particle (1) or (2); the Dirac operator $\Lambda(\mathcal L)$
corresponds to the Lorentz boost transformation $K_{i(0)}=\mathcal
L^{-1}K_i$, which places the deuteron at rest. The
$np$ amplitudes $\bar\chi^{(0)}_{Sm_S}$ and
$\bar{\chi}_{Sm_S}^{(t)}$ and the deuteron vertex function
$\Gamma_{m_d}$ are $4\times4$ matrices in spinor space.

Next, the partial-wave decompositions of Eq.~\eqref{MPWA}
and~\eqref{MFSI} have to be carried out. Since in
Ref.~\cite{KS,KS1} that has been done in great detail for the
invariant transition amplitude in PWA, here we present only the
final result for the invariant transition amplitude with FSI
included
\begin{eqnarray}
\mathcal{M}_{fi,IA}^{FSI}
&=&
\sum_{J=0,1}\sum_{S=0,1}\sum_{L^{\prime\prime}=0,2}
\sum_{LL^{\prime}\:m\mu}
i^L(-1)^{S}C_{LmS\mu}^{Jm+\mu}Y_{Lm}(\thpp,0)
\nonumber\\
&&\times\int\limits_{-\infty}^{+\infty}\! dp_0
\int\limits_0^{+\infty}\!\bp^2\,
d|\bp|\int\limits_{-1}^{+1}d\cos\theta_{\bp}\,\,
\dfrac{t^*_{LL^{\prime}}(0,|\bhp|;p_0,|\bp|;s)}{\left(\sqrt{s}/2
-E_{\bp}\right)^2-p_0^2-i0}~~
\nonumber\\&&
\times
\dfrac{\phi_{L^{\prime\prime}}\left(\mathcal{L}^{-1}(p_0-\omega/2),
\left|\mathcal{L}^{-1}(\bp-\mathbf{q}/2)\right|\right)}
{(P_f/2-q+p)^2-m^2+i0}\,
\Upsilon_{L^{\prime}L^{\prime\prime}}^\mu(\Hat\bp,\bp;q)
\label{final FSI},
\end{eqnarray}
where $\phi_{L^{\prime\prime}}(p_0,|\bp|)$ is the radial part of the deuteron
vertex function, $\Upsilon_{L^{\prime}L^{\prime\prime}}^\mu(\Hat\bp,\bp;q)$
are the
matrix elements of the Dirac operators $\Lambda(\mathcal L)$ and
the $\gamma NN$ vertex $\Gamma^{(1,2)}(q)$ between the
spin-angular momentum functions $\bar{\mathcal Y}_{L^{\prime}}$ and
$\mathcal Y_{L^{\prime\prime}}$. We calculate
$\Upsilon_{L^{\prime}L^{\prime\prime}}^\mu(\Hat\bp,\bp;q)$ using the algebra
manipulation package REDUCE. What then remains to be done is a
three-dimentional integration over $p_0$, $|\bp|$ and
$\cos\theta_{\bp}$. The integration over $p_0$ is performed
analytically with the help of Cauchy'a theorem, giving special
attention to the complicated singularity structure of the
integrand in Eq.~\eqref{final FSI}.  The remaining two-dimensional
integration is done numerically, using the programming language
FORTRAN.

\section{Results and discussions}\indent

We should mention that we can demonstrate the importance of
final-state interaction from the threshold of deuteron breakup to
the laboratory photon energy $E_\gamma\cong50$~MeV. This is
because the rank-one separable potential does not provide
satisfactory fits to the $NN$ scattering experimental data for the
spin-triplet ${}^3P^+_{0,1}$-channels above
$T_{\text{lab}}\geqslant 100$~MeV. Fortunately, it is sufficient
to take into account final states only with $J\leqslant 1$ within
this energy range.

In Fig.~1,~2 we show the results of our investigation. The curves
labeled by PWA and NR PWA depict the results of calculations in
the plane wave approximation with one-body current within the
relativistic and nonrelativistic models. Notations FSI and NR FSI
correspond to calculations in impulse approximation with
final-state interaction included within the same models.

In Fig.~1, the differential cross section
$d\sigma/d\Omega_{\Hat\bp}$, v.s. the Eq.~\eqref{c00}, at three
different laboratory energies 3, 20 and 50~MeV, in units of
microbarns, is shown as a function of the c.m. proton angle. The
sensitivity of the angular distributions to the various
final-state interaction is illustrated by a computation of
contributions from separate partial-wave channels of the final
$np$ system with $J\leqslant1$. The solid curve in Fig.~1
represents the result of the plane-wave impulse approximation.
Immediately, we can see that the relative value and sign of
final-state contributions change with increasing photon energy
$E_\gamma$. The tensor force in the initial bound and
final-scattering ${}^3S_1^+$--${~}^3D_1^+$ states is made
$P_d=4$~\%. On the other hand, transitions to the isovector final
states $^1S_0^{+}$ and $^3P_1^{+}$ play a predictably significant
role even at small $E_{\gamma}$, and their contributions become
relatively large as the photon energy increases. The contribution
of the isoscalar spin-singlet final state $^1P_1^{+}$ is
negligible for the considered range of the photon energies,
because this partial-wave amplitude is proportional to small
factor $\tfrac{\omega\mu_s}{2m}$ with $\mu_s=-0.12$ and
relativistic effects are very small. The isoscalar spin-triplet
final state ${}^3S_1^+$--${~}^3D_1^+$ becomes pronounced at about
$E_{\gamma}=50$~MeV. It is confirmed by numerical results,
obtained in Ref.~\cite{HLW}, in the framework of the
nonrelativistic theory. Moreover, in view of the interference
between isovector and isoscalar spin-triplet final states, the
contribution from the transition into ${}^3S_1^+$--${~}^3D_1^+$
final state does not vanish as $E_\gamma$ increases. On other
hand, $^1P_1^{+}$-contribution is indeed small for photon energies
up to 300~MeV. An interesting question here is the role played by
the relativistic corrections.

Fig.~2 helps to gain some insight into the relative importance of
the relativistic effects in the considered final $np$ scattering
states. We make additional calculations of the differential cross
section in the nonrelativistic phenomenological treatment of
deuteron photodisintegration: (1) using the radial wave function
of the deuteron for the nonrelativistic Graz-II separable kernel
with the deuteron $D$-state probability $P_d=4$~\%~\cite{RT}; and
(2) computing the relativistic matrix elements $t_{LL'}(0,|\bhp|;
p_0',|\bpp|;s)$ at $p_0'=0$. This provides us an opportunity to
make term-by-term comparisons of various final-state
contributions.

Our analysis shows that contribution of final-state interaction is
changed by several percent at $E_\gamma = 5$ MeV and
up to twenty percent at $E_\gamma = 50$ MeV,
when computed in the framework of the
relativistic model. We conclude that the relativistic effects
stemming from the final-state contributions should be incorporated
in the development of a rigorous theory of deuteron
photodisintegration  at higher energies. Unfortunately, our
results cannot be compared to the experimental data. The
differential cross section is much too low because we neglect the
two-body exchange currents and negative-energy components
of the bound-state vertex function and the scattering $T$ matrix.

{\bfseries Acknowlegments.}
One of the authors (K.Yu.K) greatly appreciates the help of Mr. N.
Thacker in preparation of the English version of this manuscript.
The work was supported in part by the Russian Foundation for Basic Research,
grant {No}.02-02-16542.

\newpage

\begin{center}
\includegraphics[height=68mm]{fsi_3mev_all.eps}
\vskip 12mm
\includegraphics[height=68mm]{fsi_20mev_all.eps}
\vskip 12mm
\includegraphics[height=68mm]{fsi_50mev_all.eps}
\\
Figure 1. Contribution of the separate partial-wave channels
of the final $np$ pair into the angular distribution of
c.m. differential cross section at different laboratory
photon energies.
\end{center}

\newpage

\begin{center}
\includegraphics[height=77mm]{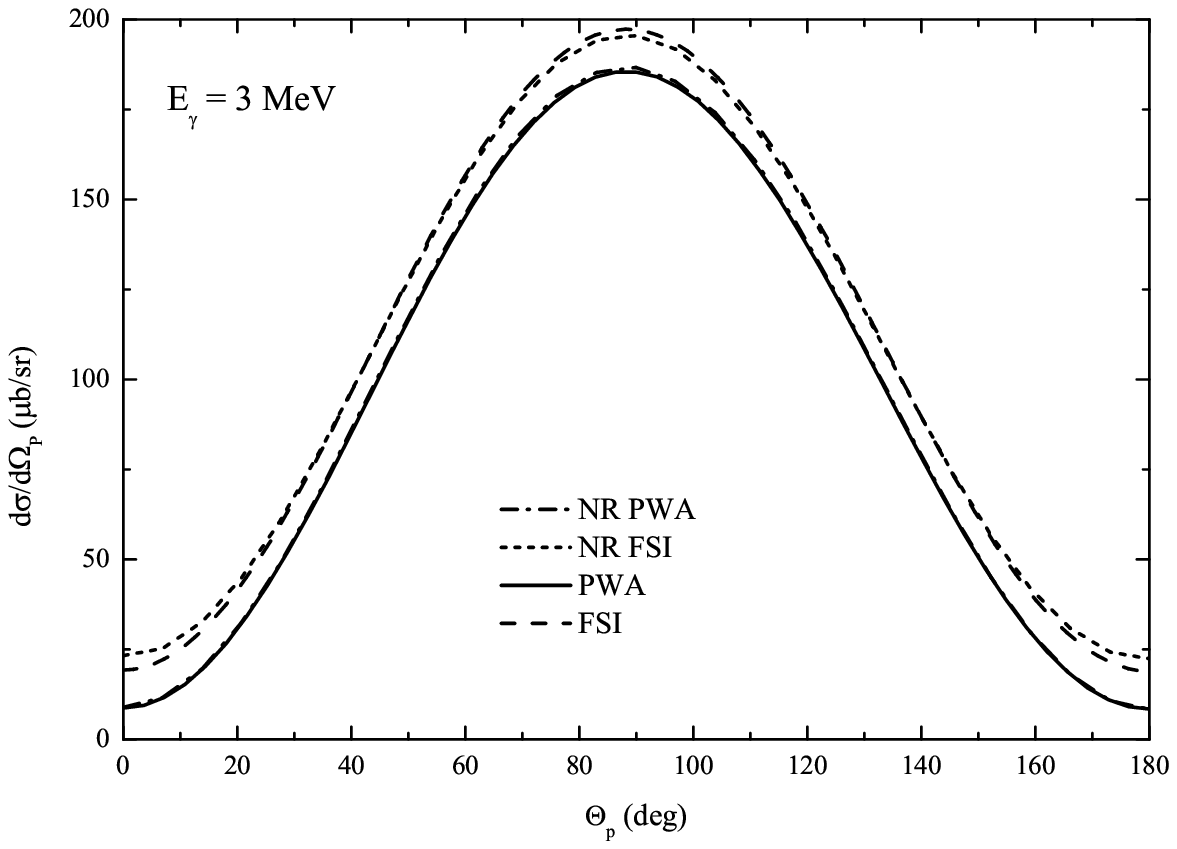}
\includegraphics[height=77mm]{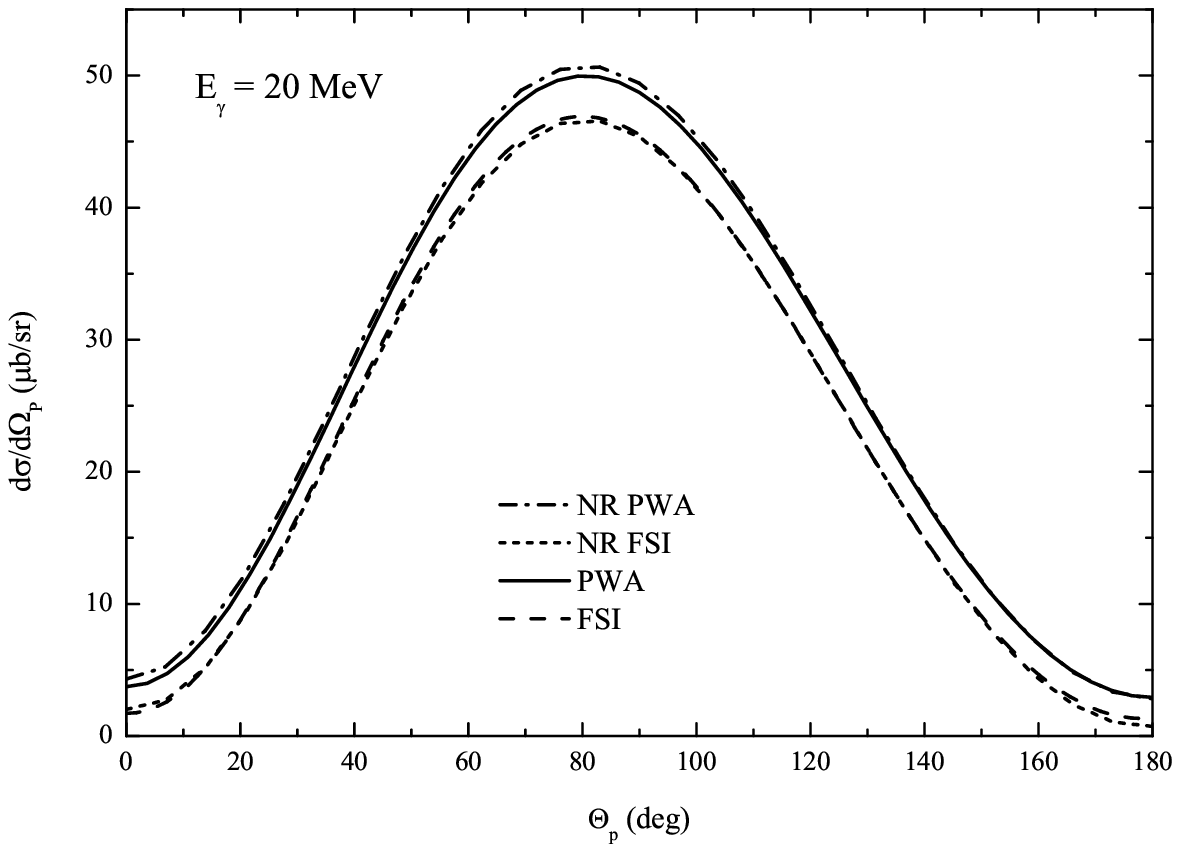}
\includegraphics[height=77mm]{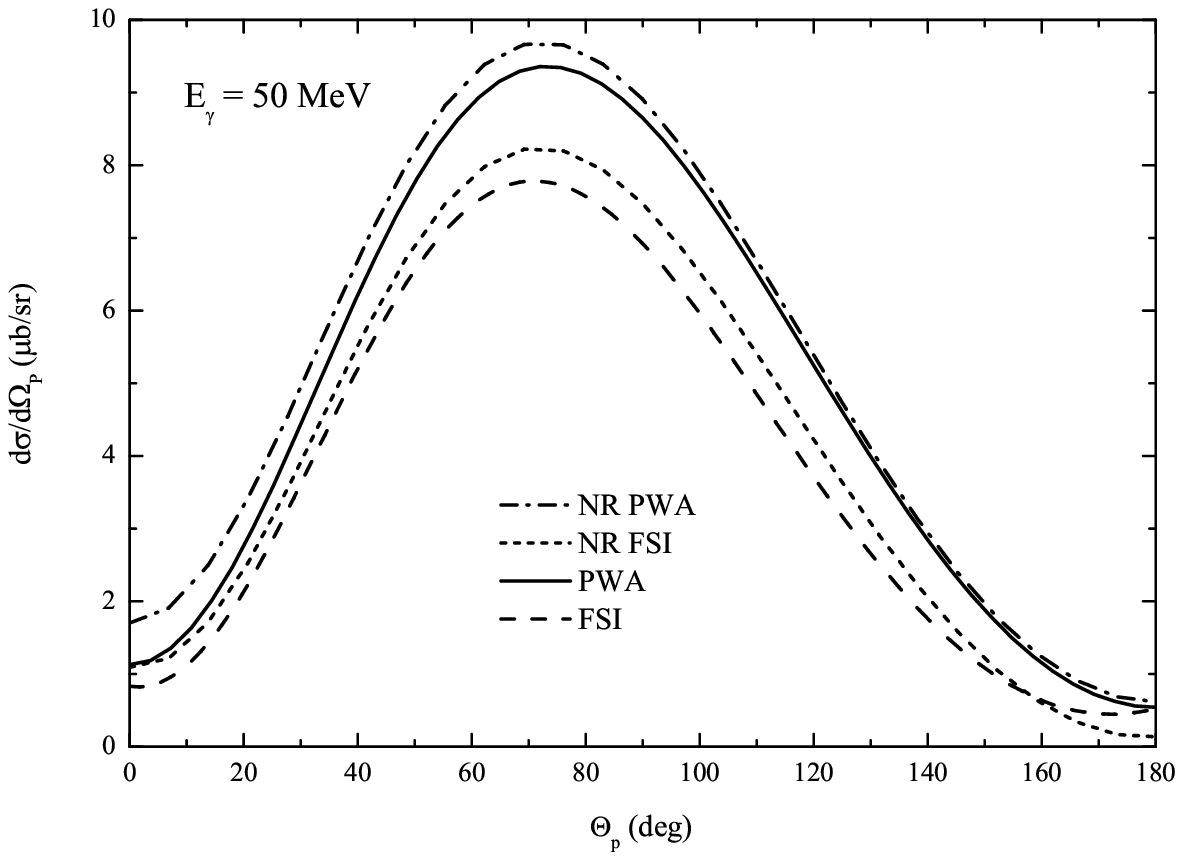}
\\
Figure 2. C.m. differential cross section at different
laboratory photon energies.
\end{center}

\end{document}